\documentclass[twocolumn,twocolappendix,apj]{aastex63_edited}

\usepackage{amssymb,amsmath,verbatim,mathtools,needspace,enumitem,graphicx,physics,microtype,mathrsfs,amssymb,xcolor,afterpage,orcidlink}

\renewcommand{\citet}[1]{\citeauthor{#1}~\citep{#1}}

\newcommand{\new}[1]{\textcolor{black}{#1}}

\def\d{{\rm d}}

\begin{document}

\title{Distinguishing the origin of eccentric black-hole mergers \\ 
with gravitational-wave spin measurements}

\author{Jakob Stegmann$\,$\orcidlink{0000-0003-2340-8140}}
\email{jstegmann@mpa-garching.mpg.de}
\affiliation{%
Max Planck Institute for Astrophysics, Karl-Schwarzschild-Str.~1, 85748 Garching, Germany
}%

\author{Davide Gerosa$\,$\orcidlink{0000-0002-0933-3579}}
\affiliation{Dipartimento di Fisica ``G. Occhialini'', Universit\`a degli Studi di Milano-Bicocca, Piazza della Scienza 3, 20126 Milano, Italy}
\affiliation{INFN, Sezione di Milano-Bicocca, Piazza della Scienza 3, 20126 Milano, Italy}

\author{Isobel Romero-Shaw$\,$\orcidlink{0000-0002-4181-8090}}
\affiliation{Department of Applied Mathematics and Theoretical Physics, Cambridge CB3 0WA, United Kingdom}
\affiliation{Kavli Institute for Cosmology Cambridge, Madingley Road Cambridge CB3 0HA, United Kingdom}
\affiliation{H.H. Wills Physics Laboratory, Tyndall Avenue, Bristol BS8 1TL, United Kingdom}

\author{Giulia Fumagalli} 
\affiliation{Dipartimento di Fisica ``G. Occhialini'', Universit\`a degli Studi di Milano-Bicocca, Piazza della Scienza 3, 20126 Milano, Italy}
\affiliation{INFN, Sezione di Milano-Bicocca, Piazza della Scienza 3, 20126 Milano, Italy}

\author{Hiromichi Tagawa$\,$\orcidlink{0000-0002-5674-0644}}
\affiliation{Shanghai Astronomical Observatory, Shanghai, 200030, People's Republic of China}

\author{Lorenz Zwick}
\affiliation{Niels Bohr International Academy, The Niels Bohr Institute, Blegdamsvej 17, DK-2100, Copenhagen, Denmark}

\begin{abstract}
It remains an open question whether the binary black hole mergers observed with gravitational-wave detectors originate from the evolution of isolated massive binary stars or were dynamically driven by perturbations from the environment. Recent evidence for non-zero orbital eccentricity in a handful of events is seen as support for a non-negligible fraction of the population experiencing external driving of the merger. However, it is unclear from \textit{which} formation channel eccentric binary black-hole mergers would originate: dense star clusters, hierarchical field triples, active galactic nuclei, or wide binaries in the Galaxy could all be culprits. Here, we investigate whether the spin properties of eccentric mergers could be used to break this degeneracy. Using the fact that different formation channels are predicted to either produce eccentric mergers with mutually aligned or randomly oriented black-hole spins, we investigate how many confident detections would be needed in order for the two models to be statistically distinguishable. If a few percent of binary black hole mergers retain measurable eccentricity in the bandwidth of ground-based detectors, we report \new{a $\sim9\,\%$ chance that we could confidently distinguish both models (Bayes factor $\ln\mathcal{B}>3$) after the fifth observing run of the LIGO-Virgo-KAGRA detector network, $\sim63\,\%$ for LIGO A\#, and $\sim98\,\%$ for the Einstein Telescope and Cosmic Explorer.}
\vspace{1cm}\end{abstract}

\section{Introduction}\label{sec:intro}
Ten years after the first direct detection of gravitational waves (GWs) from merging binary black holes (BBHs)~\citep{GW150914}, about a hundred merger events have been observed~\citep{2023PhRvX..13a1048A}. However, the astrophysical mechanism forming them remains unclear. Multiple formation channels have been proposed, which can be broadly divided into two categories. On the one hand, isolated massive binary stars may develop merging BBHs after they have interacted through a stable mass-transfer episode~\citep{vandenHeuvel2017,Inayoshi2017,Bavera2021,Marchant2021,Gallegos-Garcia2021,Olejak2021,2021PhRvD.103f3007S,vanSon2022}, in a common-envelope phase~\citep{Postnov2014,2016ApJ...819..108B,Eldridge2016,Lipunov2017,Giacobbo2018a}, or in a chemically homogeneous evolution~\citep{deMink2016,MandeldeMink2016,duBuisson2020,Riley2021}. On the other hand, BBHs may be driven to merge by some perturbation from the environment, e.g., distant tertiary companions in hierarchical triples~\citep{Blaes2002,Thompson2011,Antonini2012,Antognini2014,Antonini2014,Prodan2015,Silsbee2017,Antonini2018,Rodriguez2018,Grishin2018,Hoang2018,Stegmann2022b,Stegmann2022,2025arXiv250317006V,Stegmann2025}, scatterings in dense star clusters~\citep{Samsing2014,Antonini2016,2018PhRvD..98l3005R,Zevin2021,Mapelli2021,Chattopadhyay2023,DallAmico2024}, fly-bys and the tidal field of the host galaxies~\citep{Michaely2019,Raveh2022,2024ApJ...972L..19S}, or gaseous embeddings in active galactic nuclei (AGNs)~\citep{Goldreich2002,Bellovary2016,Fragione2019,Tagawa2020,Grishin2024,Vaccaro2024,Gilbaum2025}.

A powerful smoking-gun signature of the dynamical or otherwise externally perturbed formation of merging binaries is the presence of non-zero orbital eccentricity in a detected GW signal. This is because GW emission leads to extremely efficient angular momentum loss which rapidly circularises the orbit of a binary~\citep{Peters1964}. Thus, any system that retains a substantial residual eccentricity upon entering the sensitive frequency range of a ground-based GW detector must have formed with a much higher eccentricity. A fraction of BBH mergers with such large eccentricities is only expected if they are driven by an environmental perturbation. Conversely, the presence of non-zero eccentricity at the minimum sensitive frequency of ground-based detectors is wholly inconsistent with the expected evolution of isolated massive binary stars \citep[e.g.,][]{Belczynski2002}. Due to these considerations, the detection of even a few sources with non-zero eccentricity (or the lack thereof; \citealt{Zevin2021}) would act as a ``tip of the iceberg'' and could be used to put constraints on the fraction of mergers formed from dynamical channels in the entire observed population \citep[cf.,][]{2020PhRvD.102d3002B}.

\begin{figure*}[ht]
    \centering
    \includegraphics[width=0.9\textwidth]{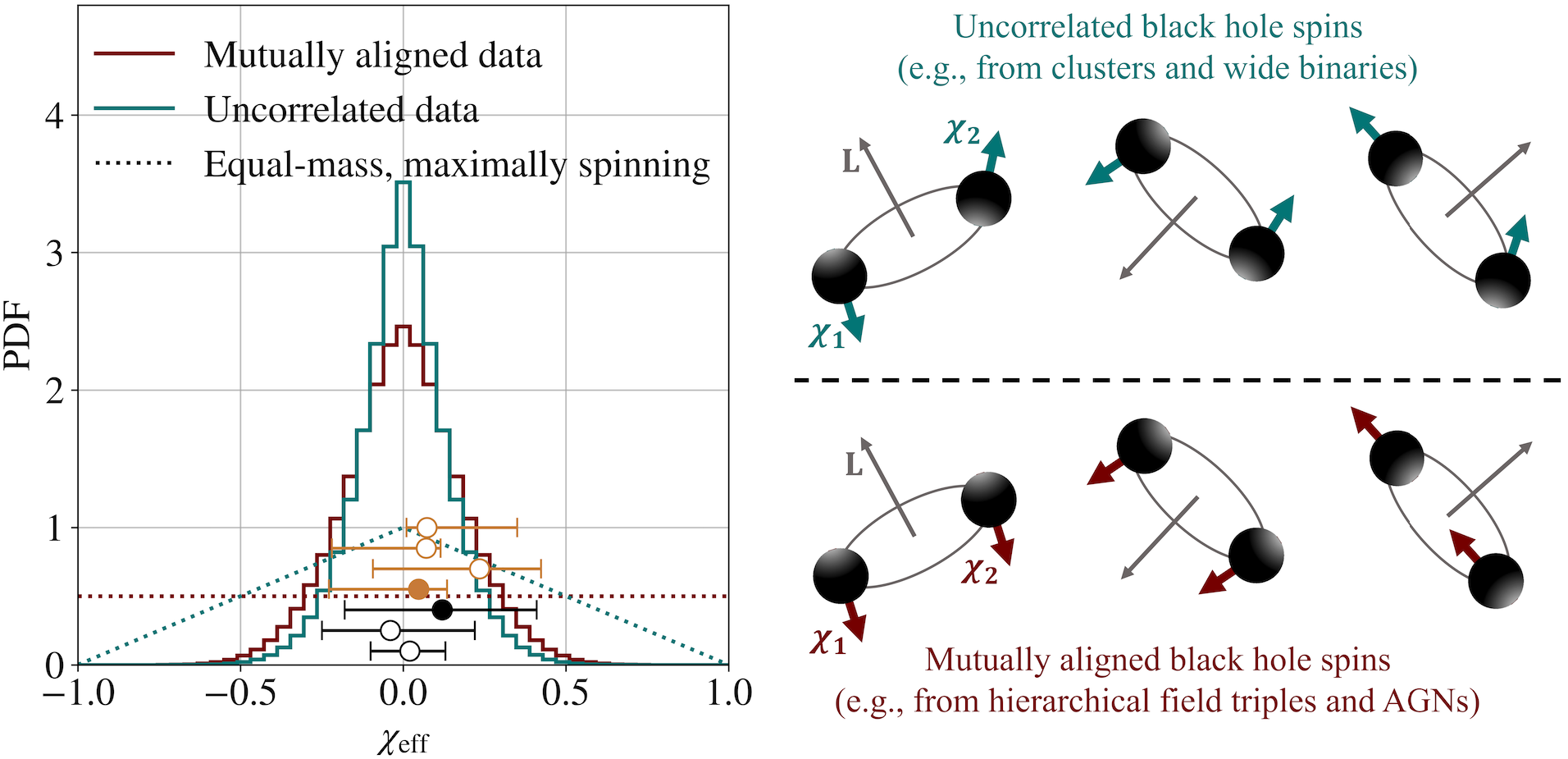}
    \caption{\textit{Left panel:} Distribution of $\chi_{\rm eff}$ for mutually aligned spin directions \new{(maroon)} and random spin directions (teal). Dotted lines indicate equal-mass ($q=1$) and maximally spinning BBHs ($\chi_1=\chi_2=1$), solid lines assume an extended mass ratio distribution and a spin magnitude distribution that peaks at lower values, which are consistent with previous GW detections (see text). Black markers show three eccentric BBH merger candidates identified by~\cite{2024arXiv240414286G} (from bottom to top: GW200129, GW190701, and GW200208\_22) and orange markers show four candidates identified by~\cite{2022ApJ...940..171R} (from bottom to top: GW200208\_22, GW190521, GW190620, and GW191109). The commonly identified candidate GW200208\_22 is highlighted by a filled marker \citep{2025arXiv250617105R}. \textit{Right panel:} Sketch of uncorrelated (top) and mutually aligned black-hole spins of eccentric mergers (bottom).}
    \label{fig:1}
\end{figure*}

Tentative evidence for compact object mergers with non-zero eccentricity has recently been found in a handful of BBH merger events \cite[e.g.,][]{2022ApJ...940..171R,2024arXiv240414286G,2024ApJ...972...65I, 2025arXiv250415833D}, though degeneracies with misaligned spin are, at present, hard to eradicate \cite[e.g.,][]{2023MNRAS.519.5352R} (although, see e.g. \citealt{2024PhRvD.109d3037D, 2025arXiv250315393M}). If this evidence becomes more robust, it would suggest that a somewhat large fraction of the BBH mergers detected thus far formed through a dynamical formation channel. Here, we consider a scenario in which additional eccentric BBH merger events are confidently detected, and investigate whether measuring their spin properties would further constrain \textit{which} of the channels that produce eccentric mergers contribute. We focus on the spin properties because, for the subset of eccentric mergers, different formation channels make very distinct predictions. The two dimensionless spin vectors $\boldsymbol{\chi}_1$ and $\boldsymbol{\chi}_2$ of the black holes and their orbital angular momentum vector $\mathbf{L}$ are either expected to point in three random directions, or the spins are mutually closely aligned ($\boldsymbol{\hat{\chi}}_1\approx\boldsymbol{\hat{\chi}}_2$) but randomly oriented with respect to $\mathbf{\hat{L}}$, as argued below and illustrated in the right panel of Fig.~\ref{fig:1}.

\begin{enumerate}
\item\label{item-cluster}
Three random directions of $\boldsymbol{\chi}_1$, $\boldsymbol{\chi}_2$, and $\mathbf{L}$ are a natural property of eccentric BBH mergers that form in dense star clusters. There, BBHs that retain measurable eccentricity in the bandwidth of current GW detectors merge during resonant three- and four-body encounters that typically involve chaotic exchanges between binary members and scatterings that heavily perturb their orbits~\citep{Samsing2014,Antonini2016,2018PhRvD..98l3005R,Zevin2021,Chattopadhyay2023}. In these chaotic encounters, merging BBHs are randomly assembled without any correlations between the spins and orbital angular momentum. As a result, the directions of $\boldsymbol{\chi}_1$, $\boldsymbol{\chi}_2$, and $\mathbf{L}$ are independent of one another and effectively randomised.

\item\label{item-ZKL}
In contrast, closely aligned spins $\boldsymbol{\hat{\chi}}_1\approx\boldsymbol{\hat{\chi}}_2$ and a random direction of $\mathbf{{L}}$ are expected for eccentric BBH mergers that result from the evolution of hierarchical triples. In these systems, the BBHs are formed from massive binary stars whose spins are likely (closely) aligned to one another by stellar processes. The gravitational perturbation from a tertiary companion promotes a subsequent merger by causing large-amplitude von Zeipel-Kozai-Lidov (ZKL)~\citep{Zeipel1910,Kozai1962,Lidov1962} oscillations of the BBH eccentricity and the orientation of its orbital angular momentum. BBHs that merge with a residual eccentricity undergo exceptionally strong ZKL oscillations while the spin evolution is highly non-adiabatic, i.e., the orbital angular momentum vector oscillates several orders of magnitude quicker than the spins could follow it through relativistic spin-orbit coupling~\citep{Antonini2018,Rodriguez2018,Liu2018,Liu2019b}. As a result, the eccentric BBH mergers retain the mutually aligned spins they inherited from their binary progenitor stars while the direction of their orbital angular momentum vector is effectively randomised. 
\end{enumerate}

Other dynamical channels, such as BBHs merging due to the tidal field of the host galaxy and stellar fly-bys, as well as BBH mergers in AGNs, have been less studied in the context of spins in eccentric events. 
However, it is plausible to expect that they also fall into either one of the aforementioned categories. 
 
First, BBHs that merge due to the perturbation from the galactic tidal field and fly-bys need to be exceptionally wide with semi-major axis $\gtrsim10^3\,\rm AU$ \citep[][]{2024ApJ...972L..19S}. Wide binaries are thought to form from initially unbound black holes which randomly pair up during the dissolution phase of young star clusters \citep{Kouwenhoven2010}, leading to uncorrelated directions of two spins and the orbital angular momentum vectors. Therefore, any eccentric BBH merger which originates from the evolution of wide massive binary stars would have three random directions of $\boldsymbol{\chi}_1$, $\boldsymbol{\chi}_2$, and $\mathbf{L}$.

In AGN disks, the spins of BBHs are expected to be mutually aligned through gas accretion due to the Bardeen-Petterson effect~\citep{Bardeen1975}, in which the alignment of the black-hole spin with a circum-black-hole disk proceeds much faster than the evolution of the spin magnitude (see \citealt{2020MNRAS.494.1203M,2023PhRvD.108h3033S} for related modeling).
However, 
the directions of the orbital angular momentum can be misaligned with the black-hole spin directions depending on the frequency of binary-single interactions. This frequency is influenced by the properties of the AGN disks, especially the 
merging locations within the AGN disks. Recent models predict frequent interactions with random angular momentum directions~\citep{Tagawa2020_spin}. In these cases, the spin-orbit alignment for the AGN channel is predicted to be similar to that of the three-body channel, and highly eccentric mergers are expected for a fraction of mergers. 
On the other hand, for the AGN channel, hierarchical mergers are inevitable~\citep{Yang2019,Tagawa2021_gap,2021NatAs...5..749G}, resulting in high masses and spins for these black holes, which can help distinguish them from triple systems.

In this work, we study how many confident detections of eccentric BBH mergers we would need before the two spin-orbit configurations of (i) three randomly distributed directions of $\boldsymbol{\chi}_1$, $\boldsymbol{\chi}_2$, and $\mathbf{L}$ and (ii) two randomly distributed directions of $\boldsymbol{\chi}_1$ and $\mathbf{L}$ and $\boldsymbol{\hat{\chi}}_1\approx\boldsymbol{\hat{\chi}}_2$ become statistically distinguishable. \new{We highlight that focusing on spin-eccentricity correlations is particularly advantageous compared to previous approaches to distinguish formation channels from the marginalised spin distribution of all mergers \citep[e.g.,][]{Rodriguez2018,Rodriguez2019,Baibhav2024}. First, eccentric mergers can \textit{only} be explained by a reduced number of formation channels, as any residual eccentricity close to merger is  inconsistent with isolated binary formation scenarios \citep{2024PhRvD.110f3012F}. Instead, eccentric mergers \textit{must} have formed through a formation channel involving dynamical driving. Second, models of such formation channels only predict the clear distinction in spin distributions~\ref{item-cluster} and~\ref{item-ZKL} for eccentric events. The majority of events that merge on effectively circular orbits can originate from a wider variety of channels and have a less certain spin distribution. For instance, a significant fraction of all mergers that form in triples is expected to have a spin-orbit precession timescale comparable or shorter than the ZKL timescale, which leads to a more diverse merger distribution of spin-orbit angles \citep{Liu2018,Rodriguez2018}. Meanwhile, an uncertain fraction of non-eccentric mergers in clusters may originate from primordial stellar binaries that experience no or only few dynamical interactions \citep{Barber2024} and whose spin directions do not fully randomise as in \ref{item-cluster}. Thus, while any viable formation channel must be consistent with the entire joint spin-eccentricity distribution, this work shows that its detectably-eccentric sub-population is particularly insightful to constrain merger origins.}

\section{Methods} \label{sec: methods}

We consider externally-driven BBH mergers that {retain some measurable eccentricity} upon entering ground-based GW detectors, which we refer to as ``eccentric BBHs''. It is typically assumed that orbital eccentricities above a threshold of several $0.01$ at a nominal GW frequency of $10\,\rm Hz$ could be measurable for events detected by the LIGO-Virgo-KAGRA (LVK) detector network \citep{Lower2018}.

Dynamical formation channels that produce eccentric BBHs make distinct predictions about the relative directions of their orbital angular momentum vector $\mathbf{L}$ and the dimensionless spin vectors $\boldsymbol{\chi}_1$ and $\boldsymbol{\chi}_2$. Whereas in some models (e.g., dense star clusters) eccentric BBHs are formed with $\mathbf{L}$, $\boldsymbol{\chi}_1$, and $\boldsymbol{\chi}_2$ pointing in three independent random directions, in other models (e.g., hierarchical field triples) only $\mathbf{L}$ is randomised but $\boldsymbol{\chi}_1$ and $\boldsymbol{\chi}_2$ are (nearly) aligned. We stress that this is only true for the subset of eccentric mergers, whereas the entire population of mergers from a particular channel can follow more complicated distributions of the spins and the orbital angular momentum. These two predictions impact the effective spin parameter \citep{2001PhRvD..64l4013D}
\begin{equation}\label{eq:chi-eff}
    \chi_{\rm eff}=\frac{\chi_1\cos\theta_1+q\chi_2\cos\theta_2}{1+q},
\end{equation}
where $0<q\leq1$ is the mass ratio between the two black holes, $\chi_1$  ($\chi_2$) is the spin magnitude of the heavier (lighter) black hole, and $\cos\theta_1=\boldsymbol{\hat{\chi}}_1\cdot\mathbf{\hat{L}}$ and $\cos\theta_2=\boldsymbol{\hat{\chi}}_2\cdot\mathbf{\hat{L}}$ are the cosines of the spin-orbit tilts angles. We focus on the effective spin parameter because it is usually the best-constrained spin parameter in GW measurements. Moreover, orbit-averaging the equations of motion of the BBHs $\chi_{\rm eff}$ remains conserved during their inspiral~\citep{2008PhRvD..78d4021R,2023PhRvD.108b4042G}, i.e., its value is the same at BBH formation and merger. While the orbit-averaged approximation technically breaks down for highly eccentric mergers \citep{2025arXiv250206952F}, we demonstrate in Appendix~\ref{sec:appendix} that resulting variations of $\chi_{\rm eff}$ are expected to be irrelevant for this work.  A concise investigation of the evolution of the individual tilt angles $\theta_{1,2}$ is presented in Appendix~\ref{costiltsev}.

Random directions of $\mathbf{L}$, $\boldsymbol{\chi}_1$, and $\boldsymbol{\chi}_2$ correspond to uncorrelated distributions $\cos\theta_1\sim\mathcal{U}(-1,1)$ and $\cos\theta_2\sim\mathcal{U}(-1,1)$. In contrast, random directions of $\mathbf{L}$ but mutually aligned spins mean $\cos\theta_1=\cos\theta_2 \sim\mathcal{U}(-1,1)$. We refer to these models as ``uncorrelated'' and ``mutually aligned'' respectively. The difference between the distribution of $\chi_{\rm eff}$ in both models is maximised if we consider the unrealistic but instructive case of equal-mass ($q=1$) and maximally spinning  ($\chi_1=\chi_2=1$) BBHs. This is shown in the left panel of Fig.~\ref{fig:1} (dotted lines), which illustrates that the mutually aligned model tends to produce more extreme values near $\chi_{\rm eff}\approx\pm1$ whereas the uncorrelated  model is more centred around zero. Both models resemble one another more closely if we consider more realistic, non-unity mass ratios and spin magnitudes.
In Fig.~\ref{fig:1}, this is shown with solid lines where we draw the mass ratio from a power-law distribution $p(q)\sim q^\alpha$ between 0.1 and 1.0 and the spin magnitudes from some Beta distribution $p(\chi_{1,2})\sim\rm Beta(a,b)$. We choose fiducial values $\alpha=1.1$, $a=1.4$, and $b=3.6$ which are consistent with the distributions inferred from the GWTC-3~\citep{2023PhRvX..13a1048A}. As a result, the mutually aligned model retains more pronounced tails of the distribution, whereas the uncorrelated model is more concentrated around $\chi_{\rm eff}=0$. Yet, the two distributions are more similar than in the previous case where $q=\chi_1=\chi_2=1$ due to a weaker contribution of the second term on the right-hand side of Eq.~\eqref{eq:chi-eff}.

Any attempt to statistically distinguish between both distributions with the current sample of eccentric candidate events (colored markers in Fig.~\ref{fig:1}) would be 
preposterous, not only because of the small number statistics but also because they were identified using eccentric aligned-spin waveform models, which may result in systematic biases against highly misaligned spins. However, we can estimate whether we could statistically prefer one model over the other with the growing data obtained from future GW observations and improved waveform modelling that accounts for precessing spins in eccentric events.

For this purpose, we assume a total number of $N_{\rm tot}$ detected BBH mergers, which we set to $2,100$ as a reference for the 5\textsuperscript{th} observing run of LVK (LVK O5) \citep{Broekgaarden2024}, $10^4$ for $A\#$~\citep{2020LRR....23....3A}, and $10^5$ for the Einstein Telescope~\citep{2025arXiv250312263A} and Cosmic Explorer (ET/CE)~\citep{Reitze2019}. We assume that a fraction $0\leq\xi_{\rm ecc}\leq1$ of them will retain \new{confidently} measurable eccentricity in the detector sensitivity window, resulting in $N_{\rm ecc}=N_{\rm tot}\times\xi_{\rm ecc}$ eccentric BBHs (rounded to the next integer). \new{We only consider events in $N_{\rm ecc}$ with sufficient evidence for eccentricity, i.e., assume a suitable selection function was applied which discards events with marginal evidence to avoid spurious contamination by circular events \citep{10.1093/mnras/stz896,Roy2025}. At present, constructing a suitable selection function that would only collect confident detections of eccentricity is difficult given that current eccentric waveform models disagree on eccentricity in most events \citep{2022ApJ...940..171R,Gupte2024}. Here, we defer such task to future work and rely on the number $N_{\rm ecc}$ of confident eccentric detections that could be obtained in that way.} For each detector configuration, we repeat the following steps $10^4$ times:

\begin{enumerate}
    \item We produce a synthetic merger population by sampling $N_{\rm ecc}$ eccentric BBHs with mass ratios drawn from a power-law distribution $p(q)\sim q^\alpha$ between 0.1 and 1.0 and spin magnitudes independently drawn from some Beta distribution $p(\chi_{1,2})\sim\rm Beta(a,b)$. We assume that the distribution parameters are drawn from $\alpha\sim\mathcal{N}(1.1,1.0)$, $a\sim\mathcal{N}(1.4,0.1)$, and $b\sim\mathcal{N}(3.6,0.1)$. The assumed uncertainty is based on the current spin magnitude distribution of GWTC-3~\citep{2023PhRvX..13a1048A}, but we note it will likely improve in the future as more events are being detected. For each of the eccentric BBH mergers we compute $\chi_{\rm eff}$ using Eq.~\eqref{eq:chi-eff} assuming either mutually aligned or uncorrelated spin directions. Lastly, we approximate the posterior distribution of each $\chi_{\rm eff}$ measurement with a truncated normal distribution between $-1$ and $1$, where the adopted standard deviation $\sigma_{\chi_{\rm eff}}$ represents typical measurement uncertainties.\label{step-1}
    \item We construct mutually aligned and uncorrelated  models that are tested against the synthetic GW data constructed in step~\ref{step-1}. To this end, we repeat step~\ref{step-1} $N\gg N_{\rm ecc}$ times with a different set of parameters drawn from $\alpha\sim\mathcal{N}(1.1,1.0)$, $a\sim\mathcal{N}(1.4,0.1)$, and $b\sim\mathcal{N}(3.6,0.1)$. These values are considered as a proxy for the best-fit values that can be obtained from the distribution of all mergers $N_{\rm tot}$.
    \label{step-2}
    \item Having constructed a synthetic population of eccentric mergers in step~\ref{step-1} and uncorrelated and mutually aligned models based on spin magnitude and mass ratio samples obtained from all detected mergers in step~\ref{step-2}, we compute Bayes factors
    \begin{equation}
        \mathcal{B}=\frac{p(\text{Data}|\text{Model})}{p(\text{Data}|\text{Alternative Model})},
    \end{equation}
    where 
    \begin{equation}\label{eq:prod}
        p(\text{Data}|\text{Model})=\prod_{i=1}^{N_{\rm ecc}}p_{\rm Model}(|\chi_{{\rm eff},i}|).
    \end{equation}
    Here, $p_{\rm Model}$ is a linear interpolation of the normalised distribution of $|\chi_{{\rm eff}}|$ obtained in step~\ref{step-2} and $\chi_{{\rm eff},i}$ are the effective spin parameter values of the eccentric BBH mergers synthesised in step~\ref{step-1}. In Eq.~\eqref{eq:prod}, we consider distributions of the absolute value of $\chi_{\rm eff}$ because our models for $\chi_{\rm eff}$ are symmetric around zero (cf.~Fig.~\ref{fig:1}). \label{step-3}
\end{enumerate}
Repeating steps~\ref{step-1} to~\ref{step-3} we obtain Bayes factor distributions for 
\begin{equation}
        \mathcal{B}=\frac{p(\text{Mutually aligned data}|\text{Mutually aligned model})}{p(\text{Mutually aligned data}|\text{Uncorrelated data})},
\end{equation}
and 
\begin{equation}
    \mathcal{B}=\frac{p(\text{Uncorrelated data}|\text{Uncorrelated model})}{p(\text{Uncorrelated data}|\text{Mutually aligned model})},
\end{equation}
to quantify the confidence to recover the correct model, i.e., the mutually aligned model in the presence of mutually aligned data and the uncorrelated model in the presence of uncorrelated data.
In the discussion below, we follow \new{\cite{Kass} and consider $2\ln\mathcal{B}\gtrsim6$ (i.e., $\log_{10}\mathcal{B}\gtrsim1.3$) to indicate strong evidence for the correct model, $-6\lesssim2\ln\mathcal{B}\lesssim6$ ($-1.3\lesssim\log_{10}\mathcal{B}\lesssim1.3$) no particular preference for either model, and $2\ln\mathcal{B}\lesssim-6$ ($\log_{10}\mathcal{B}\lesssim-1.3$)) strong evidence for the incorrect model.}

If not stated otherwise, we assume an eccentric merger fraction of $\xi_{\rm ecc}=0.05$ which agrees with predictions from dynamical channels for current ground-based detectors \citep[e.g.,][]{Antonini2014,Zevin2021}, and is consistent with the order of magnitude of eccentric candidates in the current population of GW events \citep[][]{2022ApJ...940..171R,2024arXiv240414286G}. However, it likely underestimates the fraction of eccentric mergers detectable with the next-generation detectors ET/CE, as they will be sensitive to smaller eccentricities \citep{Dorozsmai_in_prep}. Below, we also study different values for $\xi_{\rm ecc}$. For the measurement uncertainty of $\chi_{\rm eff}$, we assume as a default $\sigma_{\chi_{\rm eff}}=0.2$ for LVK O5 and A\# which is the typical uncertainty of the (quasi-circular) events identified in GWTC-3. We assume that ET/CE will allow measurements that are more precise by one order of magnitude, $\sigma_{\chi_{\rm eff}}=0.02$. Since it is unclear at what precision next-generation detectors will measure ${\chi_{\rm eff}}$ of eccentric events, we also study different values $\sigma_{\chi_{\rm eff}}$. 

We note that step~\ref{step-1} implicitly assumes that eccentric and non-eccentric mergers share the same distribution of black-hole spin magnitudes, modelled as a Beta distribution. This assumption is justified because black-hole spin magnitudes are primarily determined by the physics of stellar core collapse, which is largely independent of the environment in which the binary forms. \new{Different spin magnitudes may only arise} from hierarchical mergers in dense stellar clusters, which can produce highly spinning black holes with $\chi_i \simeq 0.7$ \citep{2021NatAs...5..749G}. However, the contribution of such ``second-generation" mergers \new{to the eccentric subset as well as all mergers} is expected to be small \citep[e.g.,][]{Rodriguez2019,Mapelli2021}.

It could also be possible that multiple formation channels contribute to the total merger rate. This would only significantly impact our analysis if these channels contributed comparably and had substantially different spin magnitude distributions. Such a scenario would require a degree of fine-tuning. In contrast, the eccentric merger fractions $\xi_{\rm ecc}$ considered in this work are consistent with all mergers being dominated by dynamical formation processes. In these channels, only a subset of binaries retains measurable eccentricity within the sensitivity range of our detectors, while the rest formed through the same mechanism but circularized before entering the detectable eccentricity regime. Moreover, any alternative formation channel would need to produce a markedly different spin distribution, which is unlikely given that spin magnitudes are predominantly governed by core-collapse physics.

\section{Results}
\begin{figure*}[ht]
    \centering
    \includegraphics[width=1\linewidth]{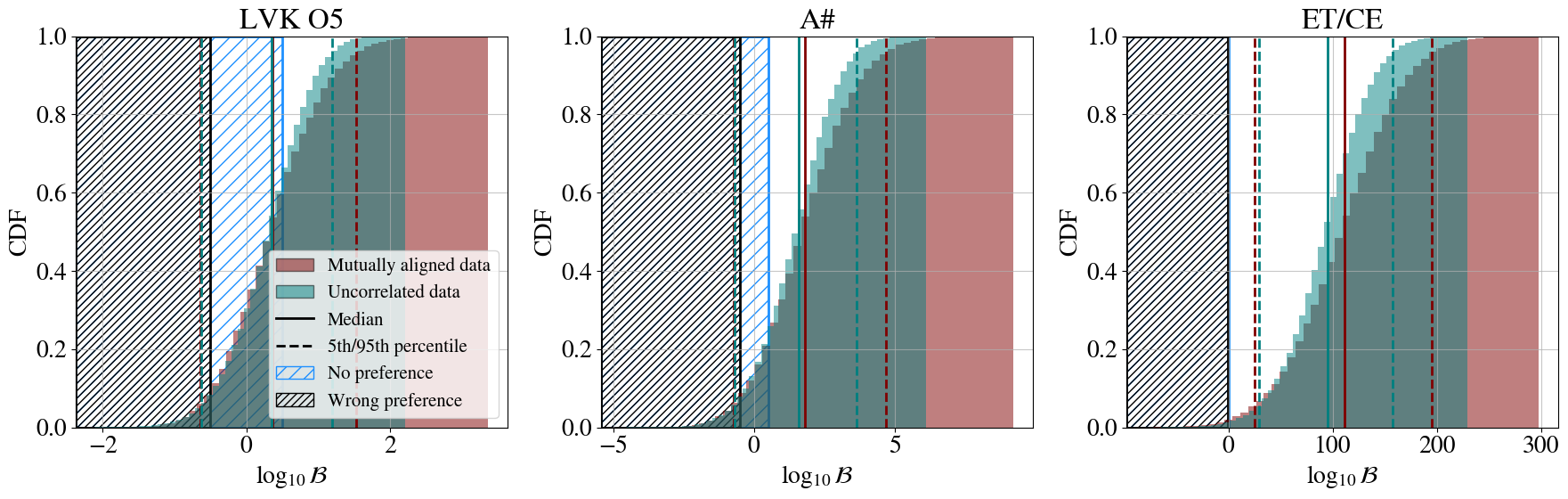}
    \caption{Bayes factor distributions that compare the mutually aligned and uncorrelated model. The \new{maroon} histogram assumes GW detections of systems with mutually aligned spins and indicates the confidence for preferring the aligned model over the uncorrelated model. Conversely, the teal histogram assumes GW detections of systems with uncorrelated spins and indicates the confidence for preferring the uncorrelated spin model over the mutually aligned spin model. Thus, in either cases a large Bayes factor \new{($\log_{10}\mathcal{B}\gtrsim1.3$)} indicates that we would correctly infer the spin properties of the underlying population. Solid and dashed vertical lines indicate the medians and 5\textsuperscript{th}/95\textsuperscript{th} percentiles of the distributions, respectively. We assume an eccentric BBH fraction of $\xi_{\rm ecc}=0.05$, a measurement uncertainty for $\chi_{\rm eff}$ of $\sigma_{\chi_{\rm eff}}=0.2$ for LVK O5 and A\# and $\sigma_{\chi_{\rm eff}}=0.02$ for ET/CE, and vary between each panel the total number of detections between $2,100$ (LVK O5), $10^4$ (A\#), and $10^5$ (ET/CE).}
    \label{fig:2}
\end{figure*}
Figure~\ref{fig:2} shows the Bayes factor distributions for the three detector configurations LVK O5 (right panel), A\# (middle panel), and ET/CE (right panel), for our default assumptions $\xi_{\rm ecc}=0.05$ and $\sigma_{\chi_{\rm eff}}=0.2$ for LVK O5 and A\# and $\sigma_{\chi_{\rm eff}}=0.02$ for ET/CE. If the eccentric BBH mergers form with mutually aligned spins (``Mutually aligned data'', \new{maroon} histograms), the Bayes factors for statistically preferring the mutually aligned spin model over the uncorrelated spin model range from $\log_{10}\mathcal{B}=0.37^{+1.14}_{-0.99}$ (LVK O5), $\log_{10}\mathcal{B}=1.84^{+2.77}_{-2.55}$ (A\#), to   $\log_{10}\mathcal{B}=111^{+87}_{-84}$ (ET/CE), where uncertainties refer to the 5\textsuperscript{th}/95\textsuperscript{th} percentiles. Similarly, for eccentric BBH mergers with uncorrelated spins (``Uncorrelated data'', teal histograms), we find Bayes factors for recovering the uncorrelated spin model over the mutually aligned spin model ranging from $\log_{10}\mathcal{B}=0.36^{+0.83}_{-0.97}$ (LVK O5), $\log_{10}\mathcal{B}=1.61^{+2.05}_{-2.27}$ (A\#), to   $\log_{10}\mathcal{B}=95^{+66}_{-62}$ (ET/CE). We consider \new{$\log_{10}\mathcal{B}\gtrsim1.3$} (non-hatched region) as a minimal threshold for substantial evidence to prefer the correct model. This is the case for \new{$\sim9\,\%$ of the realizations in LVK O5, $\sim63\,\%$ for A\#, and $\sim98\,\%$ for ET/CE}. Meanwhile, obtaining substantial evidence for preferring the incorrect model \new{($\log_{10}\mathcal{B}\lesssim-1.3$) happens for $\sim 1\,\%$ of the realizations in all detectors}. Thus, in all detector configurations, the probability of preferring the incorrect model 
is low. 
However, only with A\# and ET/CE do we expect a large probability to infer the correct model 
at high significance, 
and thus to be able to statistically distinguish the dynamical origin of the eccentric BBHs. For LVK O5, it is most likely that the data will show no strong preference for either model. 

For our default assumptions ($\xi_{\rm ecc}=0.05$ and $\sigma_{\chi_{\rm eff}}=0.2$ for LVK O5 and A\# and $\sigma_{\chi_{\rm eff}}=0.02$ for ET/CE), we also checked the expected statistics for the ongoing 4\textsuperscript{th} observing run of LVK. Assuming a total number of $N_{\rm tot}=500$ detections we find $\log_{10}\mathcal{B}=0.07^{+0.56}_{-0.43}$ and $\log_{10}\mathcal{B}=0.10^{+0.37}_{-0.49}$ for mutually aligned and uncorrelated data, respectively. We conclude that it is very unlikely that these models can be distinguished with O4 data.

\begin{figure*}
    \centering
    \includegraphics[width=1\linewidth]{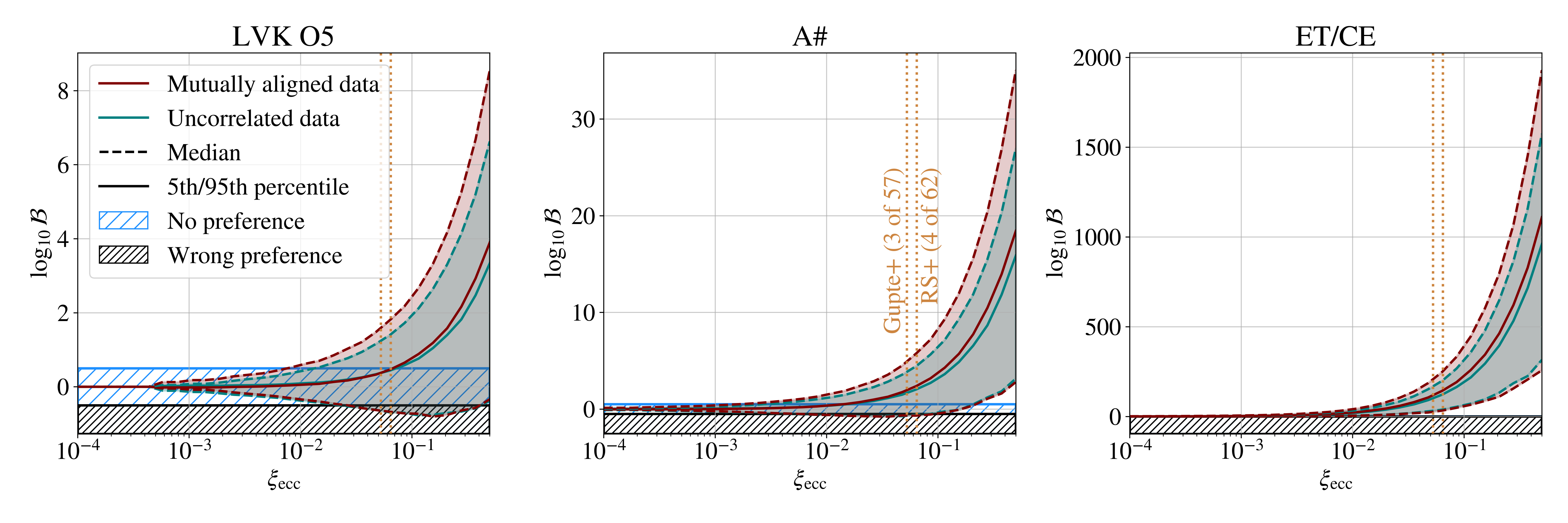}
    \caption{Percentiles of the Bayes factor distributions as a function of the eccentric BBH fraction $\xi_{\rm ecc}$. We assume that $\chi_{\rm eff}$ will be measured with an uncertainty of $\sigma_{\chi_{\rm eff}}=0.2$ for LVK O5 and A\# and $\sigma_{\chi_{\rm eff}}=0.02$ for ET/CE. As in Fig.~\ref{fig:2}, solid and dashed lines indicate the median and 5th/95th percentiles, respectively. The vertical yellow dotted lines indicate the eccentric fraction of events identified by \cite{2024arXiv240414286G} (3 of 57 analysed events; ``Gupte+'') and by \cite{2022ApJ...940..171R} (4 of 62 analysed events; ``RS+''), respectively.}
    \label{fig:3}

    \centering
    \includegraphics[width=1\linewidth]{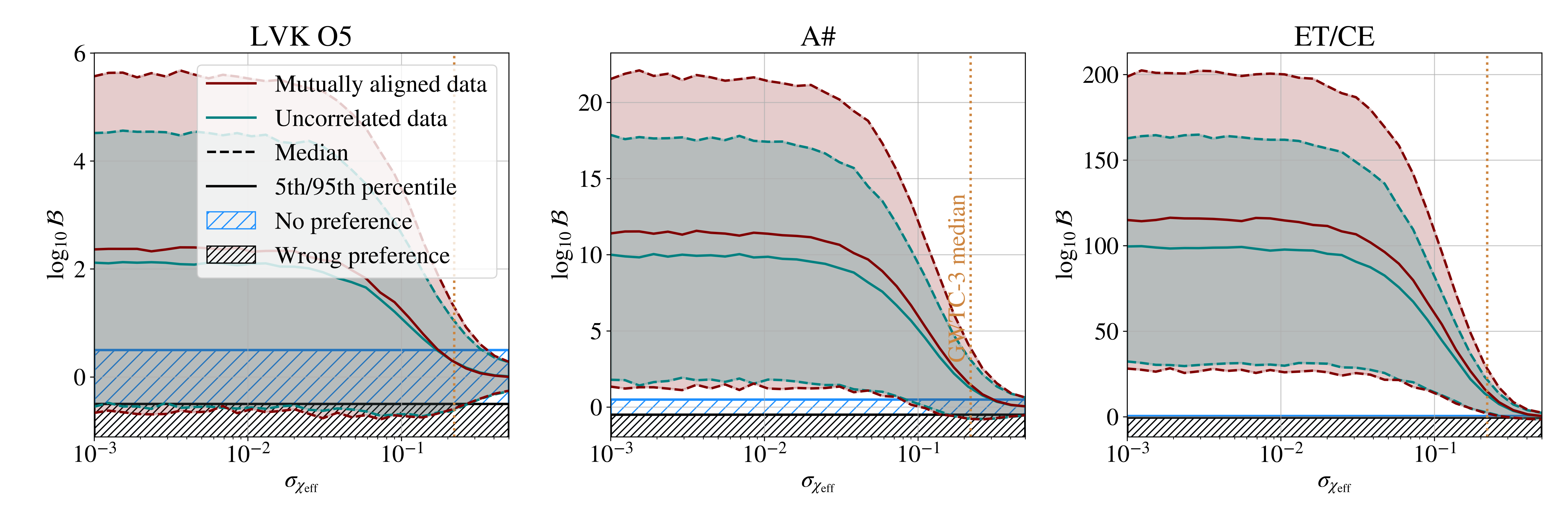}
    \caption{Percentiles of the Bayes factor distributions as a function of the eccentric BBH fraction the measurement uncertainty $\sigma_{\chi_{\rm eff}}$ of $\chi_{\rm eff}$. We assume an eccentric BBH fraction of $\xi_{\rm ecc}=0.05$. The vertical yellow dotted line indicates the typical uncertainty obtained in GWTC-3.}
    \label{fig:4}

    \centering
    \includegraphics[width=1\linewidth]{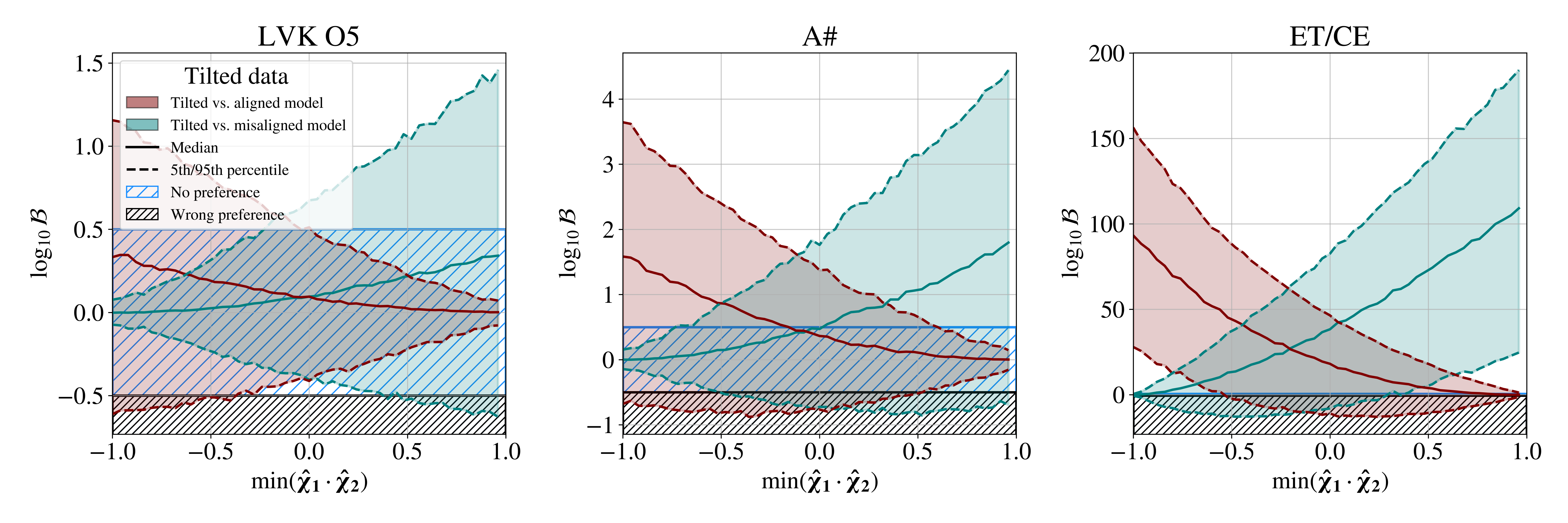}
    \caption{Percentiles of the Bayes factor distributions that compare the uncorrelated model (teal) to a tilted model (\new{maroon}), where we allow the spins to be tilted with respect to one another by $\min(\boldsymbol{\hat{\chi}}_1\cdot\boldsymbol{\hat{\chi}}_2)$. We assume an eccentric BBH fraction of $\xi_{\rm ecc}=0.05$ and a measurement uncertainty for $\chi_{\rm eff}$ of $\sigma_{\chi_{\rm eff}}=0.2$ for LVK O5 and A\# and $\sigma_{\chi_{\rm eff}}=0.02$ for ET/CE.}
    \label{fig:5}
\end{figure*}

In Fig.~\ref{fig:3}, we vary the eccentric merger fraction $\xi_{\rm ecc}$ (while keeping our default measurement uncertainty of $\sigma_{\chi_{\rm eff}}=0.2$ for LVK O5 and A\# and $\sigma_{\chi_{\rm eff}}=0.02$ for ET/CE). For LVK O5, A\#, and ET/CE, we find a minimum eccentric merger fraction of \new{$\xi_{\rm eff}\gtrsim21\,\%$, $5\,\%$, and $0.08\,\%$, respectively, is needed so that the median of the Bayes factor distributions exceeds $\log_{10}\mathcal{B}\approx1.3$}. The 5\textsuperscript{th} percentile would exceed \new{$\log_{10}\mathcal{B}\approx1.3$ if $\xi_{\rm eff}\gtrsim 37\,\%$ (A\#) and $0.6\,\%$ (ET/CE), respectively, whereas the 5\textsuperscript{th} percentile remains below $\log_{10}\mathcal{B}\approx1.3$ for LVK O5.}

In Fig.~\ref{fig:4}, we vary the measurement uncertainty $\sigma_{\chi_{\rm eff}}$ (while keeping our default eccentric merger fraction $\xi_{\rm ecc}=0.05$). For LVK O5, A\#, and ET/CE, we find that it must not be greater than 
\new{$\sigma_{\chi_{\rm eff}}\approx0.09$, $0.21$, and $0.40$, respectively, for the median to exceed $\log_{10}\mathcal{B}\approx1.3$}. The 5\textsuperscript{th} percentile would \new{$\log_{10}\mathcal{B}\approx1.3$ if $\sigma_{\chi_{\rm eff}}\lesssim 0.03$ (A\#) and $0.21$ (ET/CE)}, respectively. \new{Thus, we also repeated our entire analysis by conservatively assuming that the measurement uncertainty $\sigma_{\chi_{\rm eff}}$ of eccentric sources with ET/CE does not improve but will be the same as for LVK O5 and A\#, i.e., $\sigma_{\chi_{\rm eff}}=0.2$ (instead of our default assumption of 0.02). We find that this only marginally affects our conclusions, demonstrating that it is primarily the large number of detections with ET/CE that will allow distinguishing between both models. In particular, we find that still $\sim96\,\%$ of the realizations with $\xi_{\rm ecc}=0.05$ will have $\log_{10}\mathcal{B}\gtrsim1.3$ (compared to the $\sim98\,\%$ for $\sigma_{\chi_{\rm eff}}=0.02$ and $\sim0.02\,\%$ have $\log_{10}\mathcal{B}\lesssim1.3$ (instead of $\sim0.01\,\%$).}

Above, we assumed that in the mutually aligned model the two spins are {perfectly} aligned to one another, i.e., $\boldsymbol{\hat{\chi}}_1=\boldsymbol{\hat{\chi}}_2$. Generally, mutually aligned spins can be the result of massive binary stars whose stellar spins synchronise due to tides or other stellar binary interactions and whose black hole remnants inherit the alignment before driven to an eccentric merger due to the environment (e.g., tertiary companions in hierarchical triples or AGNs). However, some misalignment might occur if, e.g., the stars are born with uncorrelated spins or misalign during a mass-transfer episode~\citep{2021PhRvD.103f3007S} but tides are too inefficient to align them, the black holes form with spins which do not inherit the orientation from the rotation axes of the progenitor stars~\citep{2024arXiv241203461B}, or supernova kicks tilt the direction of the orbital angular momentum~\citep{2000ApJ...541..319K,2018PhRvD..98h4036G,2021PhRvD.103f3032S}. 
Thus, we investigate another ``tilted model'' where, instead of setting $\boldsymbol{\hat{\chi}}_1=\boldsymbol{\hat{\chi}}_2$, we sample $\boldsymbol{\hat{\chi}}_1\cdot\boldsymbol{\hat{\chi}}_2$ uniformly between one and a given minimum value $\min(\boldsymbol{\hat{\chi}}_1\cdot\boldsymbol{\hat{\chi}}_2)\in[-1,1)$ and assume a random azimuthal angle of $\boldsymbol{\hat{\chi}}_2$ about $\boldsymbol{\hat{\chi}}_1$. In Fig.~\ref{fig:5}, we assume that the eccentric BBH mergers form with tilted spins (``Tilted data'') with some minimum value $\min(\boldsymbol{\hat{\chi}}_1\cdot\boldsymbol{\hat{\chi}}_2)$ shown on the x-axis (with our default assumptions $\xi_{\rm ecc}=0.05$ and $\sigma_{\chi_{\rm eff}}=0.2$ for LVK O5 and A\# and $\sigma_{\chi_{\rm eff}}=0.02$ for ET/CE). \new{Maroon} and teal percentiles of $\log_{10}\mathcal{B}$ show how well we could recover the tilted model over the aligned and uncorrelated models, respectively. Note that for $\min(\boldsymbol{\hat{\chi}}_1\cdot\boldsymbol{\hat{\chi}}_2)=-1$ and $+1$ the tilted model is equivalent to the uncorrelated and mutually aligned model, respectively. Thus, the test against the mutually aligned model (\new{maroon}) converges to $\log_{10}\mathcal{B}=0$ as $\min(\boldsymbol{\hat{\chi}}_1\cdot\boldsymbol{\hat{\chi}}_2)\rightarrow1$, and vice-versa the test against the uncorrelated model (teal) goes to zero as $\min(\boldsymbol{\hat{\chi}}_1\cdot\boldsymbol{\hat{\chi}}_2)\rightarrow-1$. Crucially, the tilted model will be statistically preferred against the uncorrelated model (teal) even if we allow large spin-spin misalignments, e.g., the teal median is \new{$\log_{10}\mathcal{B}\gtrsim1.3$ for $\min(\boldsymbol{\hat{\chi}}_1\cdot\boldsymbol{\hat{\chi}}_2)\gtrsim0.7$ for A\# and for $\gtrsim-0.9$ for ET/CE, respectively.} Astrophysically, that means that we would be able to distinguish formation channels in which the spin direction would fully randomise, e.g., clusters, from those that tend to align the spins, e.g., hierarchical triples, even if we allow significant tilt angles between them.

\section{Conclusion}
Identifying the formation channel of BBH mergers is a major quest in GW astronomy. While eccentricity is a smoking-gun signature of dynamical assembly or the influence of a tertiary, several channels that produce eccentric mergers may contribute to the observed merger rate. In this work, we show that spin measurements, when restricted to the subpopulation of eccentric mergers, can delineate \emph{which} of these channels is at play.

Specifically, we investigated if $\chi_{\rm eff}$ measurements of eccentric BBH mergers can be used to distinguish between models with mutually aligned and randomly oriented spins, which would help to identify the dynamical formation channel through which they must have formed. For our default assumptions, corresponding to an eccentric merger fraction $\xi_{\rm ecc}=0.05$ and measurement uncertainty $\sigma_{\chi_{\rm eff}}=0.2$ for LVK O5 and A\# and $\sigma_{\chi_{\rm eff}}=0.02$ for ET/CE, we find an \new{$\sim9\,\%$ chance to confidently distinguish both models after LVK O5, $\sim63\,\%$ for LIGO A\#, and $\sim98\,\%$ for ET and CE (Fig.~\ref{fig:2})}. These results depend significantly  on $\xi_{\rm ecc}$ and the future values of $\sigma_{\rm eff}$, both of which are uncertain, and may correlate. 
Parameter estimation studies using eccentric waveform models have shown correlations between $\chi_\mathrm{eff}$ and eccentricity, which could cause wider uncertainties on $\chi_\mathrm{eff}$ if detected eccentricity is poorly constrained~\citep{OSheaKumar2023,2024PhRvD.109d3037D}; however, we note that since future ground-based GW detectors will be able to constrain eccentricities about two orders of magnitude lower than current detectors~\citep{2024MNRAS.528..833S}, this effect is likely to be small.

Finally, we note that our analysis is simplistic in \new{four} ways. First, we have not included any non-eccentric detections to distinguish between formation channels of BBH mergers. Since 
current models predict $\xi_{\rm eff}\lesssim\mathcal{O}(0.1)$, a given number of eccentric mergers necessarily implies there is a much larger number of circularised mergers originating from the same channel. Thus, any viable formation channel must not only recover the spin-orbit properties of eccentric mergers but also be consistent with the rest of the population, where multiple channels (including isolated binary star evolution) might contribute \citep{Zevin2021b}. In turn, focussing on the eccentric mergers has the advantage that we know they must originate from a dynamical or triple channel and that these channels make distinct predictions for the spin-orbit properties. Second, we did not consider the possibility that multiple channels contribute to the formation of eccentric mergers. We expect this to influence our analysis only if they contribute at about the same order of magnitude, in which case in step~\ref{step-2} of Sec.~\ref{sec: methods} a mixture model must be constructed that contains a fraction $0<\beta<1$ of mutually aligned mergers and a fraction $1-\beta$ of uncorrelated ones. \new{Third, we approximated the posterior distribution of $\chi_{\rm eff}$ with a truncated normal distribution, but note that in reality asymmetries may arise \citep[e.g.,][]{Ng2018}. However, we do not expect this to significantly change our qualitative results. Furthermore, establishing robust correlations between measurable eccentricity and binary spin remains an open problem in the field. Our analysis is strictly valid in the high–signal-to-noise ratio (SNR) regime; should selection effects due to eccentricity-spin measurements prove to be significant, the methodology presented in this paper can be straightforwardly applied to a higher-purity subset of the catalogue, obtained by imposing a more stringent SNR threshold. While this would increase the required number of detections, it would not alter the qualitative conclusions of the approach.}



\section*{Acknowledgements}
We thank the Sexten Center for Astrophysics, where this work was kickstarted. 
We thank Johan Samsing, Ilya Mandel, Fabio Antonini, \new{and the anonymous referee for useful input and} discussions.
J.S.~acknowledges funding from NWO Vidi Grant No.~639.042.728--BinWaves. 
D.G. and G.F. are supported by
ERC Starting Grant No.~945155--GWmining, 
Cariplo Foundation Grant No.~2021-0555, 
MUR PRIN Grant No.~2022-Z9X4XS, 
MUR Grant ``Progetto Dipartimenti di Eccellenza 2023-2027'' (BiCoQ).
and the ICSC National Research Centre funded by NextGenerationEU. 
D.G. is supported by MSCA Fellowship No. 101064542–StochRewind, MSCA Fellowship No. 101149270–ProtoBH, and MUR Young Researchers Grant No. SOE2024-0000125.
I.M.R.-S.~acknowledges the support of the Herchel Smith fund. 

\appendix

\section{Evolution of $\chi_{\rm eff}$ during eccentric BBH inspirals}\label{sec:appendix}
\begin{figure}
    \centering
    \includegraphics[width=1\linewidth]{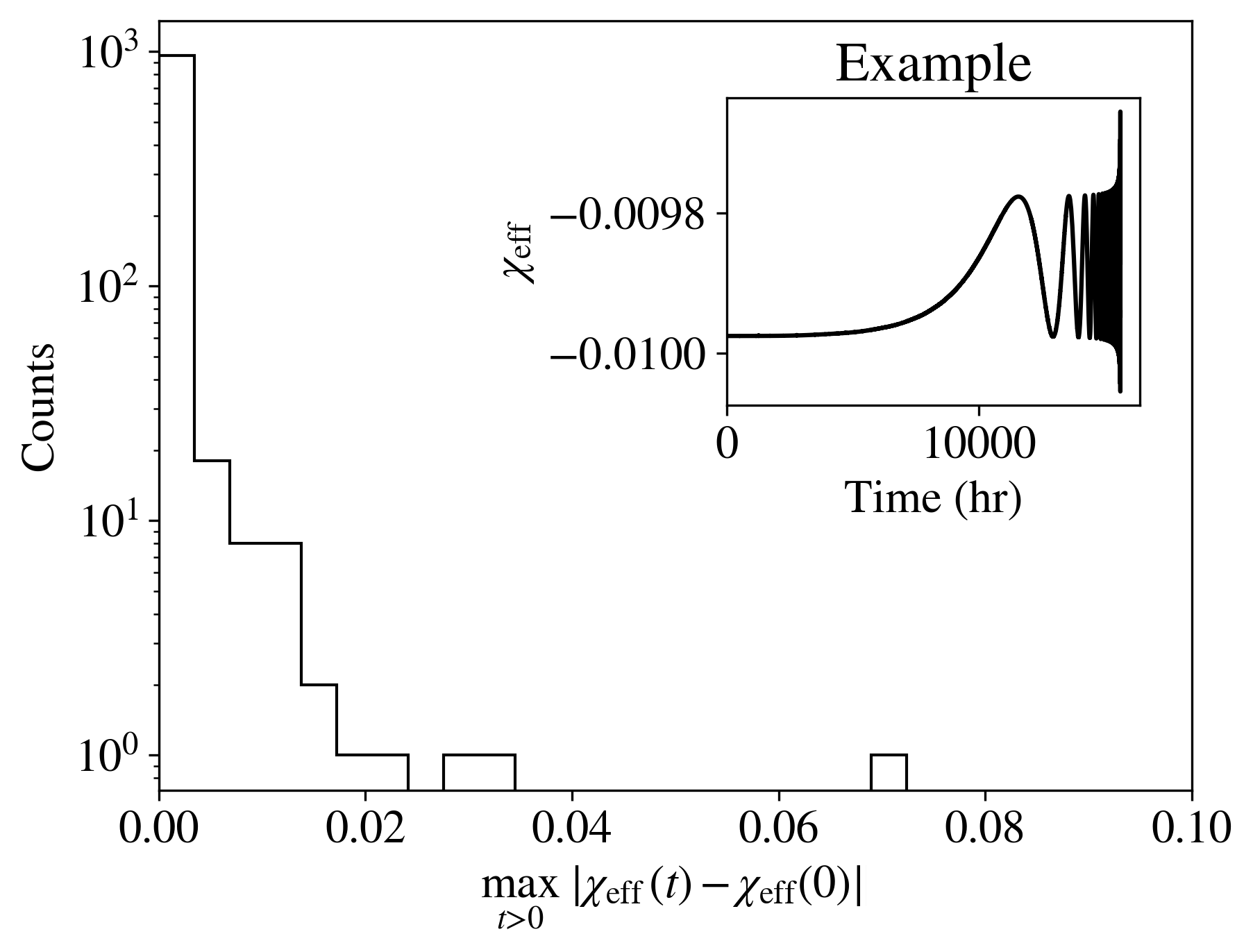}
    \caption{Distribution of the maximum difference $\max_{t>0}|\chi_{\rm eff}(t) - \chi_{\rm eff}(0)|$ of $\chi_{\rm eff}$ during the inspirals of our BBH sample with initially aligned spins and high eccentricity. The inset exemplifies the time evolution of $\chi_{\rm eff}$ of a typical system.}
    \label{fig:6}
\end{figure}

During the inspiral of a BBH, the effective spin parameter $\chi_{\rm eff}$ is a conserved quantity at second post-Newtonian (PN) order when averaging over orbital motion~\citep{2008PhRvD..78d4021R,2023PhRvD.108b4042G}. On highly eccentric orbits the orbit-averaged approximation may break down if efficient GW emission at close periapsis passages induces changes of the orbital elements on a timescale that subceeds the orbital period. As a result, the black-hole spins may precess about the orbital angular momentum vector without conserving $\chi_{\rm eff}$, so that our method to use $\chi_{\rm eff}$ at BBH formation as a proxy for $\chi_{\rm eff}$ at BBH merger would break down. In order to estimate our error, we simulate the inspiral of one thousand BBHs using the publicly available regularised direct $N$-body integrator {\tt MSTAR}~\citep{2020MNRAS.492.4131R,2023MNRAS.524.4062M} which uses the Gragg-Bulirsch-Stoer extrapolation technique~\citep{Gragg1965,Bulirsch1966} to account for the mutual gravitational interaction between the $N=2$ black holes to 3.5PN order~\citep{2004PhRvD..69j4021M}. The integration is terminated if the BBHs get as close as twelve combined Schwarzschild radii. The BBHs are initialised on a highly eccentric orbit with eccentricity $e=0.999$, total mass $m=30\,\rm M_\odot$, and periapsis $r_p=300M$ where $M=Gm/c^2$. The mass ratio of the BBHs is drawn from a power-law distribution $p(q)\sim q^\alpha$ between 0.1 and 1.0 and the spin magnitudes from some Beta distribution $p(\chi_{1,2})\sim\rm Beta(a,b)$ with our fiducial parameters $\alpha=1.1$, $a=1.4$, and $b=3.6$. The BBHs are initialised at apoapsis and their spins are assumed to be aligned. Figure~\ref{fig:6} shows that for the majority of BBH inspirals $\chi_{\rm eff}$ does not oscillate by more than $\sim0.02$ and none of them by more than $\sim0.1$ around the initial values. Based on these simulations, we conclude that our results are not significantly affected by treating $\chi_{\rm eff}$ as a conserved quantity.

\section { Evolution of $\cos \theta_1$  and  $\cos \theta_2$ during eccentric BBH inspirals}
\label{costiltsev}

Unlike the effective spin parameter $\chi_{\rm eff}$, the spin tilt angles do not remain constant during the inspiral. Instead, they precess within precession cones whose amplitudes evolve due to GW emission, while remaining effectively fixed on the shorter precession timescale.
The presence of distinct timescales in the dynamics allows for an efficient treatment of spin evolution through a double-averaging procedure: first over the orbital timescale and then over the precession timescale \citep{2015PhRvD..92f4016G, 2023PhRvD.108b4042G}. This technique has been successfully applied to study spin-precessing, quasi-circular BBHs and, more recently, eccentric systems as well \citep{2023PhRvD.108l4055F}.
Using the extended version of the \texttt{precession} code presented by \cite{2023PhRvD.108l4055F}, we investigate the evolution of the spin tilts in a population of eccentric BBHs in the mutually aligned and uncorrelated models as described in Sec.~\ref{sec:intro}. We simulate $10^5$ binaries with total mass $M = 20 M_{\odot}$; the mass ratios are drawn from a power-law distribution $p(q) \sim q^{\alpha}$ with $\alpha = 1.1$ and $q \in [0.1, 1.0]$; spin magnitudes are either sampled from a Beta distribution $p(\chi_{1,2}) \sim \text{Beta}(a, b)$ with $a = 1.4$ and $b = 3.6$ or fixed to the maximal value $\chi_1 = \chi_2 = 1$.
BBHs are assumed to form with the initial eccentricity $e = 0.999$ and semi-major axis $a \sim 1.1 \times 10^5 GM/c^2$, corresponding a GW frequency $f_{\rm GW} \sim 0.95$ Hz. We evolve the systems until they reach an eccentricity of $e = 0.1$ at $f_{\rm GW} = 10$~Hz.
\begin{figure*}
    \centering
    \includegraphics[width=0.85\textwidth]{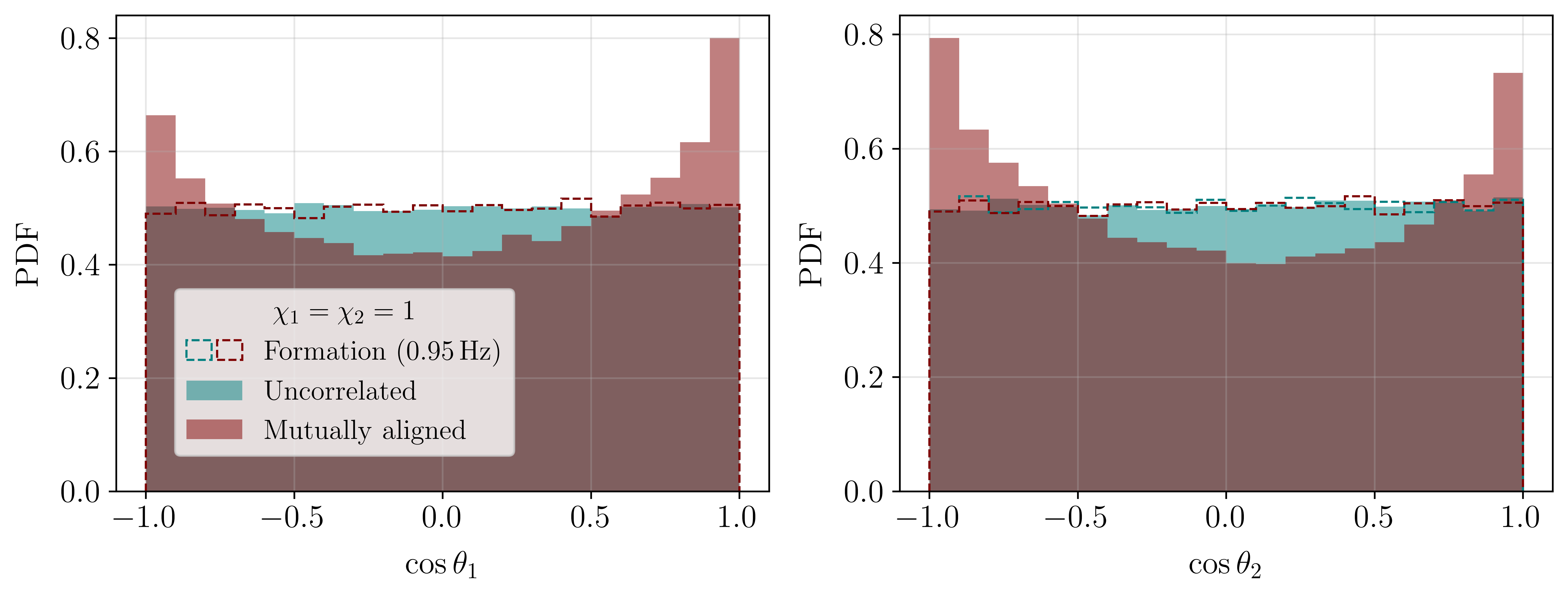}

$\,$\\

    \centering
    \includegraphics[width=0.85\textwidth]{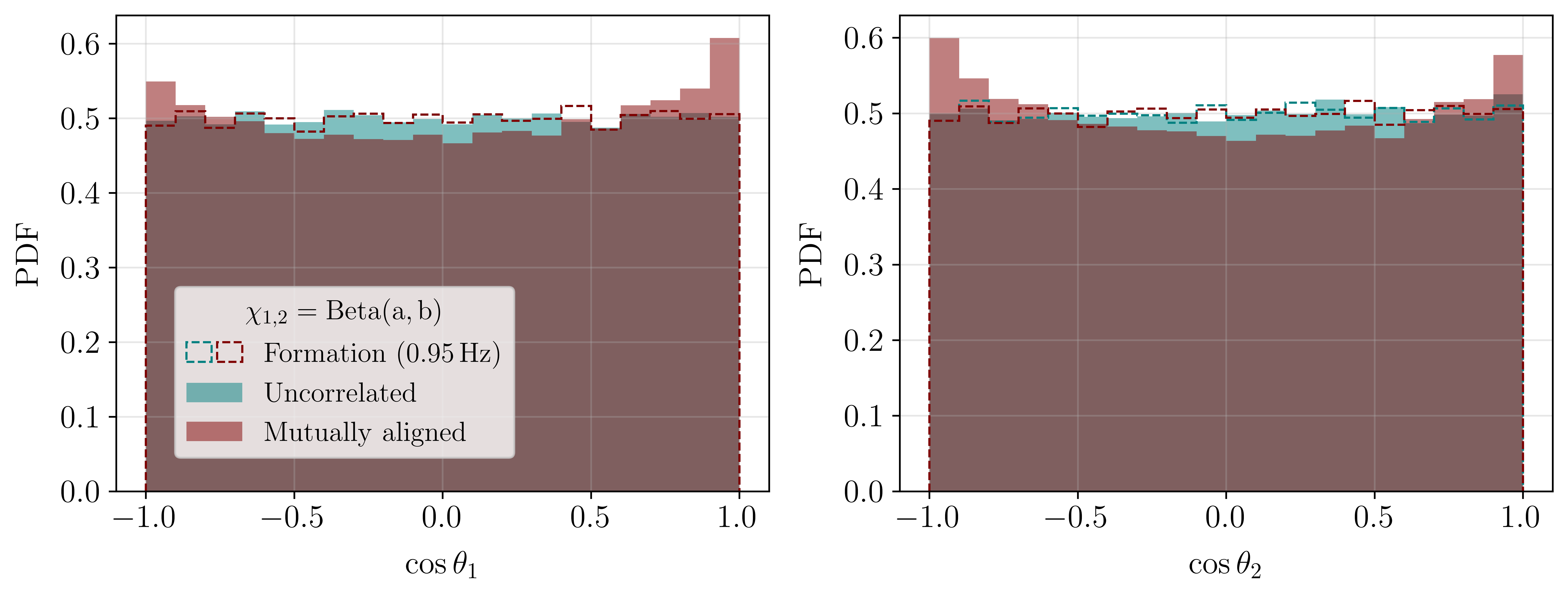}
    \caption{Distribution of the cosine of spin tilt angles, $\cos\theta_1$ (left panels) and $\cos\theta_2$ (right panels). We assume spin tilts to be either mutually aligned \new{(maroon)} or uncorrelated \new{(teal)} at formation. Distributions at detection ($f_{\rm GW} = 10$ Hz) and BBH formation ($f_{\rm GW} = 0.95$ Hz) are shown as filled and dashed histogram, respectively. We assume either maximally spinning BH (top panels) or spin magnitudes extracted from a Beta distribution as in Fig.~\ref{fig:2} (bottom panels).}
    \label{fig:7}
\end{figure*}

Figure \ref{fig:7} shows the resulting evolution of the spin tilts $\theta_{1,2}$. As previously reported in quasi-circular binaries~\citep{2007ApJ...661L.147B,2015PhRvD..92f4016G}, initially isotropic (i.e., uncorrelated) spin tilt distributions remain isotropic throughout the inspiral, even in the presence of eccentricity \citep{2023PhRvD.108l4055F}.
On the other hand, deviations appear in the case of initially aligned spin tilts. In these systems, the spin tilt distributions evolve away from  mutual alignment, exhibiting a tendency to cluster around $\cos\theta_i = \pm 1$. This behavior, already identified in the quasi-circular case~\citep{2013PhRvD..87j4028G}, is here found to persist in the eccentric regime as well. The effect is especially pronounced in the maximally spinning case, consistent with the expectation that spin-orbit and spin-spin couplings are stronger for higher spin magnitudes.

\bibliographystyle{yahapj_edited}
\bibliography{origineccspin}

\end{document}